\documentclass[a4paper]{article}

\usepackage{INTERSPEECH2020}

\usepackage{booktabs}
\usepackage{multirow}
\usepackage{nicefrac} 
\usepackage{amssymb}
\usepackage{float}
\usepackage{mathtools, nccmath}

\newcommand{\dff}{d_{\text{ff}}}

\title{Attention-based Neural Beamforming Layers for Multi-channel Speech Recognition}
\name{Bhargav Pulugundla$^{1,2}$, Yang Gao$^1$, Brian King$^1$, Gokce Keskin$^1$,  Harish Mallidi$^1$, Minhua Wu$^1$, Jasha Droppo$^1$, Roland Maas$^1$\thanks{This work was done when the first author was an intern at Amazon.}}
\address{$^1$Amazon Inc., USA\\$^2$Brno University of Technology, Speech@FIT and IT4I Center of Excellence, Czechia}
\email{bhargav.pulugundla@gmail.com, \{ygaa, bbking, gkeskin, mallidih, wuminhua, drojasha, rmaas\}@amazon.com}
\begin{document}
\maketitle
\begin{abstract}
Attention-based beamformers have recently been shown to be effective for multi-channel speech recognition. However, they are less capable at capturing local information. In this work, we propose a 2D Conv-Attention module which combines convolution neural networks with attention for beamforming. We apply self- and cross-attention to explicitly model the correlations within and between the input channels. The end-to-end 2D Conv-Attention model is compared with a multi-head self-attention and superdirective-based neural beamformers. We train and evaluate on an in-house multi-channel dataset. The results show a relative improvement of 3.8\% in WER by the proposed model over the baseline neural beamformer.  
\end{abstract}
\vspace{0.1cm}
\section{Introduction}
\label{sec:intro}
The performance of automatic speech recognition (ASR) systems has improved a lot in the recent years mostly due to the advances in deep learning~\cite{Hinton}. However, far-field ASR, with its applications such as virtual assistants, is still challenging under reverberant and noisy conditions. In practice, signal processing techniques including beamforming, denoising and dereverberation are applied to mitigate the effect of background noise, echo and speech overlap.

Typically, beamforming and acoustic modeling are trained independently as two separate modules. First, a beamformer aims to produce an enhanced single channel signal from multiple microphone inputs. This signal is then passed to the ASR model. Beamforming works by exploiting the spatial information, i.e. microphone array geometry and noise field, to combine multiple microphone inputs and amplify the signal from a specific looking direction. Some popular beamformers are trained to either maximize a signal level objective such as signal-to-noise ratio~\cite{merlsri} of the output or the Minimum Variance Distortionless Response (MVDR) criterion~\cite{mvdr}.  Beamforming is a linear system in either time or frequency domain where there is no explicit use of the correlation of multiple input channels. Furthermore, to design a beamforming system, a prior knowledge of microphone array geometry is required and different microphone array geometries usually lead to very different systems; in other words, beamforming is device-dependent. 

Some of the neural network-based beamforming techniques have overcome the previously mentioned limitations, i.e.~lack of multi-channel correlations and device dependency, while outperforming conventional methods. Various neural adaptive beamforming techniques were applied to estimate masks for speech and noise and also to predict beamforming filter coefficients~\cite{mask,merl,taranab}. Some studies have proposed training beamforming jointly with acoustic modeling~\cite{mcjournal,googlehome}. Recently, many research have paid attention to end-to-end  multi-channel speech recognition. ~\cite{Braun2018,attmultisensor} use a shared LSTM to compute the attention scores over each channel.~\cite{spatialattention} introduces spatial attention to weight multiple looking directions of a neural beamformer. An auditory attention beamformer with Connectionist Temporal Classification back-end was explored in~\cite{auditory}.

In this paper, we investigate a new attention based layer for multi-channel speech recognition that will explicitly calculate two types of correlations between the input channels through self-attention and cross-attention. We compare it to the neural beamforming and multi-head attention based approaches. We also study the impact of the different attention modules on the performance of a multi-head attention layer. We train and evaluate our models on in-house datasets consisting of anonymized utterances. The results show that the attention based models outperform the traditional neural beamforming based method and our proposed two-dimensional multi-head attention layer model yields a relative gain of 3.8\% in word error rate (WER). 

The rest of the paper is as follows. Section 3 describes the proposed architecture and different  multi-channel front-ends explored along with the back-end. The experimental setup and results are discussed in Sections 4 and 5 and Section 6 concludes the paper.
\begin{figure*}[ht!]
	\centering
	\scalebox{.6}{
	\includegraphics[]{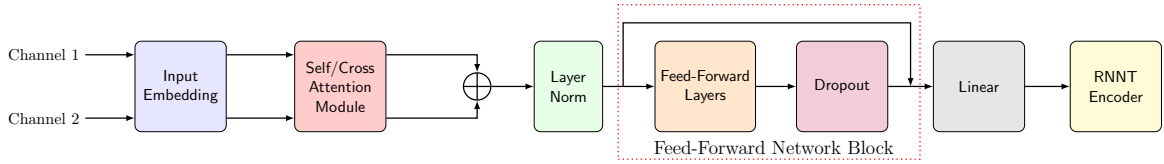}
	}
	\vspace{-0.1cm}
	\caption{Attention based multi-channel ASR}
	\label{fig:beamformer}
	\vspace{-0.3cm}
\end{figure*}
\section{Neural Beamforming}
\label{sec:baseline}
Beamforming belongs to a class of techniques that combine the signals from several sensors to emphasize a desired source while attenuating interference from other directions. This can be achieved by applying a time-invariant linear filter on the input signals. Given a microphone array with $N$ sensors at locations $m_i$ where $i = 0, 1, ...., N-1$, the final output of beamformer in frequency domain can be written as
\begin{equation}
    Y(\omega) = \boldsymbol{H}^T(\omega) \boldsymbol{F}(\omega, \boldsymbol{m}),
\end{equation}
where $\boldsymbol{H}$ are the vectors of frequency responses and $\boldsymbol{F}$ are the spectra of the signals produced by the sensors. Different type of beamformers correspond to different filters, leading to different $\boldsymbol{H}$. For example, the delay-and-sum beamformer weights the signal based on the distance from source to each sensor and sum the weighted signals. Whereas, the minimum variance distortionless response (MVDR) beamformer has the constraint that, in the absence of noise, the output of the beamformer is equivalent to the desired source signal. The well-known superdirective beamformer is a MVDR with $sinc$ diffuse noise field.

In neural network based beamforming, we treat $\boldsymbol{H}$ as learnable parameters and they can be initialized randomly or using weights from a pre-trained superdirective beamformer. During training, the parameters are updated to optimize the ASR loss function. We learn a set of $\boldsymbol{H}$ to represent multiple looking directions and then combine them linearly or through convolution. 
\section{Proposed Multi-Channel System}
We propose an attention-based end-to-end multi-channel speech recognition system. It consists of four main components.
\begin{itemize}
	\item A convolution-based input embedding module to project each input channel to a subsampled low dimensional vector. 
	\item The attention module performs beamforming on the embedded convolution features of each input channel. We experiment with different variations of  the multi-head attention and compare their performances. We describe these in Section~\ref{sec:attention_modules}. Two kinds of multi-head attentions are applied on the input channels to capture the correlations between them: \textbf{Self-attention}, with the keys, queries and values coming from the same input channel and \textbf{cross-attention}, where the key-value pair is derived from the same input channel and the query is from another channel. The outputs of attention modules on all input channels (two for each) are merged into a single channel representation. For merging, we tried concatenation but found summation to be better. The weights of this module are shared by all channels.
	\item Feed forward network block followed by a linear layer helps enhance the feature representation.
	\item The enhanced signal is then passed to an RNN Transducer for training the back-end ASR system. 
\end{itemize}  
The block diagram of the proposed architecture is shown in Figure~\ref{fig:beamformer}. 
\vspace{-0.2cm}
\subsection{Attention Modules}
\label{sec:attention_modules}
\subsubsection{Multi-Head Attention}
Multi-head attention (MHA), first introduced in~\cite{mha} for neural machine translation, enables sequence-to-sequence modelling without the use of recurrent neural networks. Given three linearly-projected $d_k$ dimensional vectors, key ($\hat{K}$), query ($\hat{Q}$) and value ($\hat{V}$), as inputs to single head attention module, the scaled-dot-product attention mechanism is computed using equation~\ref{eq:attention}.
\vspace{-0.2cm}
\begin{equation}\label{eq:attention}
   \mathrm{Attention}(\hat{Q},\hat{K},\hat{V}) = \mathrm{softmax}(\frac{\hat{Q}\hat{K}^T}{\sqrt{d_k}})\hat{V}
\end{equation}
where scaling by the square root of $d_k $ ensures that the softmax gradients do not get too small when $d_k $ is large. In the multi-head attention setup, the input vectors are split into $h$ chunks (i.e. heads), and independent attention functions are applied to each head in parallel. The attention output from the heads are then concatenated as shown in equation~\ref{eq_mha}.
\begin{equation}
\begin{aligned}\label{eq_mha}
	\mathrm{MultiHead}(\hat{Q},\hat{K},\hat{V}) &= \mathrm{Concat}(\mathrm{head_1}, ..., \mathrm{head_h})W^O\\
	\text{where}~\mathrm{head_i} &= \mathrm{Attention}(QW^Q_i, KW^K_i, VW^V_i)\\
\end{aligned}
\end{equation}
The readers can refer to~\cite{mha} for the details.
\begin{figure}[h!]
	\centering
	\scalebox{.55}{
	\includegraphics[clip]{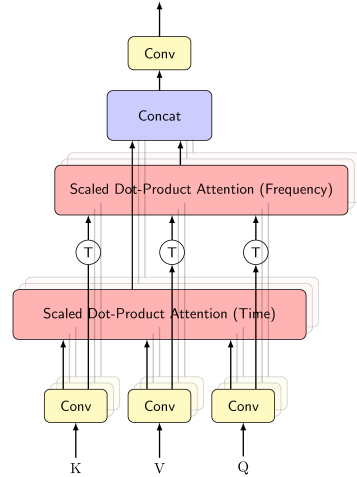}
	}
	\caption{2D Conv-Attention}
	\label{fig:2d}
	\vspace{-0.4cm}
\end{figure}

\begin{table*}[t!]
	\caption{Relative WER performance of the attention models to the neural beamformer based baseline. $d_{ff}/h/d_k$ represent the inner dimension of FFN block/number of heads or convolution channels/dimension of keys ($\hat{K}$). Positive/negative values of relative WER indicate degradation/improvement.}
	\label{tab:results}
	\begin{center}
		\vspace{-0.2cm}
    \begin{tabular}{c|cccccc|cc}
			\hline\rule{0pt}{2.0ex}
			\multirow{2}{*}{Model} & Input & Attention &
			\multirow{2}{*}{$\dff$} & \multirow{2}{*}{$h$} & 
			\multirow{2}{*}{$d_k$} & 
			train & Relative
      & params \\
			& Embedding & Dim & & & & steps & WER
      & $\times10^6$ \\
			\hline\rule{0pt}{2.0ex}
			Neural  Beamformer
			& N & - & - & -
			& - & 200K & 0.0
      & 59 \\
			\hline\rule{0pt}{2.0ex}
			Multi-Head Attention
			& Y & 128 & 1024 & 4 & 32 & 200K & -1.6
      & 59 \\
			\hline\rule{0pt}{2.0ex}
			2D Conv-Attention
			& Y & 128 & 1024 & 4 & 128 & 200K & \textbf{-3.8}
      & 60  \\
			\hline
		\end{tabular}
	\end{center}
	\vspace{-0.3cm}
\end{table*}

\subsubsection{Two-Dimensional (2D) Conv-Attention}
Although MHA models have been very successful in many speech applications, they are less capable in modeling the finer neighboring local patterns. On the other hand, convolution neural networks  (CNN) are good at learning the local information and stacking multiple CNN layers can capture the global context as well. Combining self-attention with convolution has been shown to model both the local correlations and global information effectively~\cite{conformer,qanet,conv-sa}. 

 While MHA computes attention on the time axis to model the temporal dependencies, 2D Conv-Attention explores the idea of  performing joint time-frequency analysis. We use a  combination of multi-head attention and convolution networks to model both the temporal and spectral dynamics of a speech signal~\cite{tflstm,2d}. 2D Conv-Attention employs convolution layers  to project the input $n$-channel keys, queries and values into $d_k$ dimensional vectors. We then apply scaled dot-product attention on the time and frequency axis in parallel by transposing the input vectors. Their outputs are later concatenated across the channels. Finally, it is transformed back to the original dimension through another convolution layer. Figure~\ref{fig:2d} illustrates the two-dimensional or 2D convolution multi-head attention module.
 \begin{equation}
 	\begin{aligned}
 	  \text{2D Conv-Attention} &= \mathrm{Concat}(\mathrm{channel}^t_1, ..., \mathrm{channel}^t_n,\\
     &\quad \quad  \mathrm{channel}^f_1, ..., \mathrm{channel}^f_n) * W^O
 	 \end{aligned}
  \vspace{-0.3cm}
 \end{equation}
  \begin{align*}
 	\medmath{\text{where},} &\; \medmath{\mathrm{channel}^t_i = \mathrm{Attention}(W^Q_i *Q, W^K_i*K, W^V_i*V),} \\ 
 	\medmath{\mathrm{channel}^f_i} = &\;  \medmath{\mathrm{Attention}((W^Q_i*Q)^T, (W^K_i*K)^T, (W^V_i*V)^T)^T, }\\
 	* \; \text{denotes} & \; \text{the convolution operation} \\
 	\text{and}\; \mathrm{W^O} &\; \text{represents filters of 2$n$ channels}
\end{align*}
\vspace{-0.7cm}

\subsection{RNN Transducer}
The RNN Transducer (RNNT) model, is a  streamable sequence-to-sequence model consisting of an encoder (transcription) network, a prediction network and a joint network~\cite{graves,rnnt-google}. Unlike a typical sequence-to-sequence model, RNNTs process an input sequence $X_t = (x_1,x_2,\cdots,x_t)$ of length $t$ periodically to predict a sub-word output sequence $Y_u = (y_1,y_2,\cdots,y_u)$ of length $u$. The encoder network transforms the acoustic features into higher-level representations and the prediction network produces another high-level representation conditioned on the history of non-blank label predictions from the model. The encoder and prediction network outputs are fed into a feed-forward based joint network followed by a softmax layer to produce the conditional probability distribution over the target labels. The encoder and prediction networks are analogous to the acoustic and language model of a traditional ASR system.

\section{Experimental Setup}
\label{sec:experiments}
\subsection{Data}
We train and evaluate the models on 20,000 hours of machine transcribed noisy data. The utterances are de-identified speech queries from our in-house dataset. The utterances are machine transcribed using a teacher model. The data contains real-life environmental noise and background speech. Our test set contains $45$ hours two channel audio transcribed by human.

\subsection{Architecture}
All our systems are trained on the real and imaginary short-time Fourier transformed features extracted from the input channels. They are computed using a window size of 25ms and a window shift of 10ms. We convert the complex features into rectangular coordinates. The input at time frame $t$ is stacked with 2 frames to the left i.e.~downsampled to 30ms for low frame rate modeling. The time attention is computed over three stacked frames. 

 We use RNN Transducer for the back-end. In our experiments, we stack 5 layers of uni-directional LSTMs for the encoder  and  2 uni-directional LSTMs for the decoder. Both parts have 1024 for each of the LSTM layers. A single feed forward layer with 512 hidden units  is used for the joint network. The beamformer and RNNT layers are jointly trained with the Adam optimizer using the TensorFlow toolkit. While the attention layers are Glorot normal initialized, the RNNT layers are Glorot uniform initialized. The RNNT is trained on enhanced single channel output of the attention module. 
 
\subsubsection{Baseline - Neural Beamformer}
The neural beamformer (NB) receives raw audio signals from two channels as input. We pick two microphones that are located diagonally in a 7 microphone array geometry, with 6 of them placed in an equispaced circle. The same two microphones are chosen every time. Multiple beamformers are trained with 7 looking directions and are combined using convolution~\cite{multigeometry}. We refer to this model as our baseline.

\subsubsection{Input Embedding}
The input key and query vectors are transformed into subsampled vectors using two CNN layers with 4 output channels and a stride of 2 on the frequency axis. Subsampling the input also helps fit a bigger training batch into the GPU memory. The encoded vectors of each channel are then used for beamforming. We do not apply input embedding on the baseline model since we compute energies for the looking directions from the complex features.

\subsubsection{Attention Models}
The attention models are trained on the same two-channel input as the neural beamformer. The convolution layers in the 2D Conv-Attention architecture use a kernel size of 3  and have 4 output channels. The attention outputs of the channels are merged to produce a single representation. We apply layer normalization on the single-channel output for faster convergence. It is then fed to the feed-forward network block. Similar to the position-wise feed-forward layer block of a Transformer encoder~\cite{mha}, we include a block of two linear layers with a ReLU non-linear activation between them. The inner layer has 1024  hidden units ($d_{ff}$).  The linear layers can also be substituted by two 1D-convolution layers as shown in~\cite{conformer}. The feed-forward block is highlighted in Figure~\ref{fig:beamformer}. A residual connection is added over the block followed by a linear layer. For regularization, we use a dropout rate of 0.1. Some of the other hyperparameters chosen are mentioned in Table~\ref{tab:results}. 

\section{Results} 
Table~\ref{tab:results} compares the ASR performance of the attention based models with a traditional neural beamformer. Our first set of results showed similar WER from the NB and MHA based models. The attention based models with input embedding show a minimum relative gain of 1.6\% in WER over the neural beamformer. The best performing 2D Conv-Attention yields a total relative improvement of 3.8\%. Various modules were  applied on the attention models and an ablation study was conducted on them.  The relative WER results of study with respect to the best performing 2D Conv-Attention are summarized in Table~\ref{tab:study}.

We study the importance of various modules of the multi-head attention beamformer on the overall model. First, we remove the transformation of input features with the convolution embedding module and see a significant drop of 2.3\% in the performance. A similar drop was observed with the MHA model  as seen in Table~\ref{tab:results}. Second, applying self-attention i.e.~without modeling the explicit correlations between channels, the WER increases by 1.8\%. Third, we remove the feed-forward network block and it leads to a substantial 3.1\% performance degradation. This may be because the non-linearity produces a better feature representation of the signal-channel output of the multi-channel layer. Last, computing attention only on the temporal dynamics shows a reduction of 2.3\%. Among the various modules, the feed-forward block, attention on  both time \& frequency,  and input embedding prove most beneficial.  

\begin{table}[H]
	\begin{center}
		\caption{Ablation study of the various modules of the 2D Conv-Attention model. We remove one of the modules while applying the others. For 1D Conv-Attention, we compute attention only on the time axis. Table shows the relative degradation in WER.}
		\label{tab:study}
		\begin{tabular}{l|c}
			\hline\rule{0pt}{2.0ex}
			\multirow{2}{*}{Model} & Relative   \\
			& WER \\
			\hline\rule{0pt}{2.0ex}
			2D Conv-Attention  & \textbf{0.0}  \\
			- Input Embedding & +2.3 \\
			- Cross-attention &  +1.8 \\
			- FF Network Block & +3.1 \\
			- 2D + 1D Conv-Attention & +2.3 \\
			\hline
		\end{tabular}
	\end{center}
\end{table} 
\section{Conclusions}
In this paper, we propose an end-to-end attention based multi-channel speech recognition system. Our model is invariant to the microphone geometry and also calculates the explicit correlation between input channels. We combine the attention mechanism with convolution neural networks to model the local and global dependencies. The proposed 2D Conv-Attention model shows a relative 3.8\% and 2.2\% improvement over a traditional neural beamformer and multi-head attention based model respectively, with similar number of parameters.
\bibliographystyle{IEEEtran}

\bibliography{refs}

\end{document}